
%
\input harvmac

\def\Titleh#1#2{\nopagenumbers\abstractfont\hsize=\hstitle\rightline{#1}%
\vskip .5in\centerline{\titlefont #2}\abstractfont\vskip .5in\pageno=0}
%
\def\UAa{Department of Physics and Astronomy}
\def\UAb{The University of Alabama}
\def\UAc{\it Box 870324, Tuscaloosa, AL 35487-0324, USA}
%
%
\def\hb{\hfil\break}
\def\ie{\hbox{\it i.e.}}

\def\VEV#1{\left\langle #1\right\rangle}
\catcode`\@=11 

\def\lsim{\mathrel{\mathpalette\@versim<}}
\def\gsim{\mathrel{\mathpalette\@versim>}}
\def\@versim#1#2{\vcenter{\offinterlineskip
    \ialign{$\m@th#1\hfil##\hfil$\crcr#2\crcr\sim\crcr } }}
\def\boxit#1{\vbox{\hrule\hbox{\vrule\kern3pt
      \vbox{\kern3pt#1\kern3pt}\kern3pt\vrule}\hrule}}

\def\cl{\centerline}
\def\etal{{\it et al.}}

\def\t1{{\tilde 1}}
\def\tA{{\tilde A}}
\def\tg{{\tilde g}}
\def\tp{{\tilde \gamma}}

\def\tf{{\tilde f}}
\def\tt{{\tilde t}}
\def\a3{{\alpha_3}}

\def\eV{\,{\rm eV}}
\def\MeV{\,{\rm MeV}}
\def\GeV{\,{\rm GeV}}
\def\TeV{\,{\rm TeV}}

\def\undertext#1{$\underline{\smash{\vphantom{y}\hbox{#1}}}$}
\def\NPB#1#2#3{Nucl. Phys. B {\bf#1} (19#2) #3}
\def\PLB#1#2#3{Phys. Lett. B {\bf#1} (19#2) #3}

\def\PRD#1#2#3{Phys. Rev. D {\bf#1} (19#2) #3}

\def\PRT#1#2#3{Phys. Rep. {\bf#1} (19#2) #3}

\def\UAHEP#1{University of Alabama preprint UAHEP#1}
%
%
\nref\Lou{L. Clavelli, \UAHEP{921}, (to be published).}
\nref\DataG{Particle Data Group, Review of Particle Properties, (to
appear in Phys. Rev. D).}
\nref\EK{G. Eilam and A. Khare, \PLB{134}{84}{269}.}
\nref\CER{B. Campbell, J. Ellis, and S. Rudaz, \NPB{198}{82}{1}.}
\nref\BCR{R. Barbieri, M. Caffo, and E. Remiddi, \PLB{83}{79}{345}.}
\nref\GR{T. Goldman and D. Ross, \NPB{171}{80}{273}.}
\nref\HEB{T. Hebbeker, Aachen preprint PITHA 91/17, (to appear in
the Proceedings of the LEP-HEP 91 Conference, Geneva, 1991).}
\nref\SUGA{P. Nath, R. Arnowitt and A.H. Chamseddine, {\it Applied
N=1 Supergravity} (World Scientific, Singapore 1983);\hb
H.P. Nilles, \PRT{110}{84}{1};\hb
A.B. Lahanas and D.V. Nanopoulos, \PRT{145}{87}{1}.}
\nref\HK{H.E. Haber and G.L. Kane, \PRT{117}{85}{75};\hb
R. Barbieri, Riv. Nuo. Cim. {\bf 11} (1988) 1.}
\nref\Dick{R. Arnowitt and Pran Nath, ``Testing Supergravity Grand
Unification", (to appear in the Proceedings of the LEP-HEP 91
Conference, Geneva, 1991); ``SUSY Mass Spectrum in $SU(5)$
Supergravity Grand Unification'', Texas A \& M University preprint.}
\nref\LYN{J.L. Lopez, K. Yuan, and D.V. Nanopoulos, \PLB{267}{91}{219}.}
\nref\others{J. Ellis and F. Zwirner, \NPB{338}{90}{317};\hb
M. Drees and M.M. Nojiri, \NPB{369}{92}{54};\hb
K. Inoue, M. Kawasaki, M. Yamaguchi, and T. Yanagida,
\PRD{45}{92}{328};\hb
S. Kelley, J.L. Lopez, D.V. Nanopoulos, H. Pois and K. Yuan,
\PLB{273}{91}{423}; Preprints CTP-TAMU-104/91/UAHEP9115 (to appear
in Phys. Lett. B); CERN-TH.6498/92/CTP-TAMU-16/92/UAHEP927;\hb
G.G. Ross and R.G. Roberts, Oxford preprint RAL-92-005.}
\nref\BGM{R. Barbieri, L. Girardello, and A. Masiero, \PLB{127}{83}
{429}.}
\nref\KY{K. Yuan, in preparation.}
\nref\KLN{L.E. Iba\~n\'ez and C. L\'opez, \NPB{233}{84}{511};\hb
C. Kounnas, A. Lahanas, D.V. Nanopoulos, and M. Quiros,
\NPB{236}{84}{438}.}
\nref\CHGNO{L3 Collaboration, B. Adeva \etal, \PLB{233}{89}{530};\hb
ALEPH Collaboration, D. Decamp \etal, \PLB{236}{90}{86};\hb
OPAL Collaboration, M.Z. Akrawy, \etal, \PLB{240}{90}{261}.}
\nref\HIGGS{ALEPH Collaboration, \PLB{265}{91}{475};\hb
J.L. Lopez and D.V. Nanopoulos, \PLB{266}{91}{397};\hb
V. Barger and K. Whisnant, \PRD{42}{90}{138}.}
\nref\CC{L. Clavelli, P. Coulter, B. Fenyi, C. Hester, P. Povinec, and
K. Yuan, in preparation.}
\nref\EHNOS{J. Ellis, J.S. Hagelin, D.V. Nanopoulos, K.A. Olive, and
M. Srednicki, \NPB{238}{84}{453}.}
\nref\Sciama{D.W. Sciama, Mon. Not. R. Astron. Soc. 230(1988)13.}
\nref\EOSS{J. Ellis, K.A. Olive, S. Sarkar, and D.W. Sciama,
          \PLB{215}{88}{404}.}
\nref\Ellis{J. Ellis, CERN preprint TH6193/91, (to appear in
the Proceedings of the LEP-HEP 91 Conference, Geneva, 1991).}
\nref\DEQ{S. Dawson, E. Eichten, and C. Quigg, \PRD{31}{85}{1581};\hb
UA1 Collaboration, C. Albajar \etal, \PLB{198}{87}{261}.}
\nref\AEN{I. Antoniadis, J. Ellis, and D.V. Nanopoulos,
          \PLB{262}{91}{109}.}
\nref\CUSB{CUSB Collaboration, \PLB{156}{87}{233}.}
\nref\KK{Wai-Yee Keung and A. Khare, \PRD{29}{84}{2657}.}
\nref\LK{L. Krauss, \NPB{227}{83}{556}.}
%
\nfig\I{Best fit to the 3-gluon decays of six quarkonia states assuming
a gluino mass of $400\GeV$. The experimental errors are increased
according to ``method 2''.}
\nfig\II{$\chi^2/6$ as a function of $\a3(M_Z)$ for the 3-gluon decays
of six quarkonia states treated according to method 2 but assuming a
heavy gluino (mass $400\GeV$).}
\nfig\III{$\chi^2$ contours for the fit to 5 vector quarkonia according
to ``method 1" as a function of $\a3(M_Z)$ and the gluino mass $m_\tg$.}
\nfig\IV{Fit to the 5 vector quarkonia using the ``best fit" values from
Eq.~\XIII{a}.}
\nfig\V{$\chi^2$ contours for the fit to 6 vector quarkonia according
to ``method 2" as a function of $\a3(M_Z)$ and the gluino mass $m_\tg$.}
\nfig\VI{Best fit to the 6 vector quarkonia with the expanded errors of
method 2. The fit parameters are given by the central values of
Eq.~\XIV.}
\nfig\VII{The allowed values of the gluino mass in the soft SUSY breaking
picture with $m_{1/2}=0$ are bounded by the closed figure shown here.
The range of allowed gluino masses at given ${\tA}_t$ corresponds to
taking the full allowed range of the remaining parameters $m_0$,
$m_t$, and ${\rm tan}\beta$.}
\leftline{\titlefont THE UNIVERSITY OF ALABAMA}
\Titleh{\vbox{\baselineskip12pt\hbox{UAHEP924}}}
{\vbox{\cl{Best Fit to the Gluino Mass}}}
\cl{L. CLAVELLI, P.W. COULTER, and KAJIA YUAN}
\bigskip
\bigskip
\cl{\UAa}
\cl{\UAb}
\cl{\UAc}
\bigskip
\bigskip
\bigskip
\bigskip
\cl{ABSTRACT}
\bigskip
Assuming that perturbative QCD is the dominant explanation for
the narrowness of the vector quarkonia, we perform a $\chi^2$
minimization analysis of their hadronic decays as a function of
two parameters, the mass of the gluino and the value of $\a3(M_Z)$.
A value below $1\GeV$ for the gluino mass is strongly preferred.
Consequences for SUSY breaking scenarios are discussed.
\bigskip
\Date{May, 1992}

Recently it has been pointed out that the quarkonia data can be made
consistent with minimal supersymmetric grand unification if the gluino
and photino masses are below half the $Z^0$ mass \Lou. In addition, the
quarkonia data becomes consistent with the world average
measurements of the strong
fine structure constant, $\a3(M_Z)$, if these masses are not above the
$\Upsilon$ region. The result of Ref.~\Lou\ was based on differences in
the running of the strong coupling constant in the presence of gluinos.
If, however the gluino mass is below half the $\Upsilon$ mass a second
effect comes into play. Namely, the possibility for a quarkonium state
to decay into gluino-containing final states affects the analysis of
$\a3$ at the relevant scale for that decay. In the current work we
seek a best fit value to the gluino mass, $m_\tg$, taking both
effects into account. This allows us to consider the full range of
relevant gluino masses in a $\chi^2$ analysis and to determine the
best fit value of the gluino mass $m_\tg$.

This work depends on the following basic assumption: {\it The dominant
explanation for the narrowness of the quarkonia states including the
$J/\psi$ and $\phi$ is perturbative QCD.}

Any future models for the quarkonia data that rely on very large
non-perturbative effects or relativistic corrections should be considered
as alternative to the present analysis and judged on the basis of their
relative physical plausibility. Small non-perturbative corrections, of
course, would only modify our results by small amounts.

We analyze up to second order in QCD the hadronic decay rates of six
quarkonia states, $\phi(1019)$, $J/\psi(3097)$, $J/\psi(3686)$,
$\Upsilon(9460)$, $\Upsilon(10020)$, and $\Upsilon(10350)$.
Ignoring possible higher order or non-perturbative effects and
relativistic corrections each of these defines within errors the strong
coupling constant at an appropriate scale $\mu_S(i)$
\eqn\I{\a3(\mu_S(i))=\alpha_{3,i}\pm{\delta}_i,\quad (i=1,\ldots,6).}
Assuming the gluino mass, $m_\tg$, lies above half the mass of
the quarkonium state, these values are independent of the gluino mass
and are given in Ref.~\Lou\ using the 1992 branching ratio averages of
the Particle Data Group \DataG. In the current work we consider also
lower values of $m_\tg$ so that the $\alpha_{3,i}$ become
functions of $m_\tg$.

If the gluino mass lies below half the quarkonium mass, there are
decays into two gluinos ($\tg\tg$) and into two gluinos
plus one gluon ($\tg\tg g$) which compete with the
standard three gluon ($ggg$) decay. These gluino-containing final
states \EK\ however are suppressed by four powers of
$m_q/m_{\tilde q}$ and are negligible for currently allowed values of
the squark mass $m_{\tilde q}$. The dominant gluino-related correction
to the quarkonium width is therefore the two-gluon plus two-gluino
($gg\tg\tg$) final state with no intermediate squarks.
The corresponding decay rate as a function of the gluino to quark mass
ratio $m_\tg/m_q$ was written down in Ref.~\CER\ by applying
a color+spin correction factor to the rate for the $ggq{\bar q}$
decay as calculated in Ref.~\BCR.  The result is
\eqn\II{{\Gamma(^3S_1({\bar q}q)\rightarrow gg\tg\tg)\over
         \Gamma(^3S_1({\bar q}q)\rightarrow ggg)}
   ={3\a3(\mu_S,m_\tg)\over \pi}R(r),}
where $r=2m_\tg/m(^3S_1)\simeq m_\tg/m_q$. For small $r$ (but $r>0.1$),
\eqnn\III
$$\eqalignno{R(r)=-\ln r+{9\over 32(\pi^2-9)}\Bigl[
&-3.56-r^2(8\ln r+0.95)+{28\over 27}\pi^2r^3\cr
&+r^4(-{16\over 3}\ln^2r+1.1\ln r - 8.1)+{\cal O}(r^5)\Bigr].&\III\cr}$$
Eq.~\III\ is infrared divergent as $r \to 0$. A correct treatment \EK\
shows that $R(0)\approx 1.57$. We use the above expression for $R$ in
region $r>0.1$ and use a quadratic interpolating polynomial to join $R$
smoothly to its value at $r=0$. The results presented here are
not sensitive to the exact form of the cutoff.
$R(r)$ falls rapidly for increasing $r$. For $r>0.5$ it is necessary to
use the exact integral expression for $R(r)$.
It is possible \Lou\ to choose the scale $\mu_S(i)$ so that
the known (first order) standard model corrections to the 3 gluon decay
rate vanish identically. The hadronic decay rate into states of
dissimilar quarks is then
\eqn\IV{\Gamma(^3S_1({\bar q}q)\rightarrow {\rm hadrons})=
\Gamma(^3S_1({\bar q}q)\rightarrow ggg)+
\Gamma(^3S_1({\bar q}q)\rightarrow gg\tg\tg)+\cdots}
We have then
\eqn\V{\alpha^3_{3,i}(\mu_S,m_\tg)\biggl(1+
{3\alpha_{3,i}(\mu_S,m_\tg)\over \pi}R(r)\biggr)=\alpha_{3,i}^3.}
The $gggg$ and $ggq{\bar q}$ correction to Eq.~\IV\ is absorbed into
the three gluon term by the choice of scale $\mu_S$ \Lou.
The coupling $\alpha_{3,i}$ on the
right hand side of Eq.~\V\ is the result obtained in Ref.~\Lou\ assuming
no gluino contribution ($m_\tg>m(^3S_1)/2$).

Our analysis proceeds as follows. For fixed value of the gluino mass we
solve Eq.~\V\ to obtain six values of the strong coupling constant at
six scales appropriate to each of the vector quarkonia.
The extrapolation to these scales from the $Z^0$ mass is done using the
three loop renormalization group equation,
\eqn\VI{Q{{\rm d}\a3(Q)\over {\rm d}Q}=-{{\alpha}^2_3(Q)\over 2\pi}\Bigl[
a+{b\over 4\pi}\a3(Q)+{c\over 16\pi^2}{\alpha}^2_3(Q)\Bigr],}
where
\eqna\VII
$$\eqalignno{a&=11-{2\over 3}n_f(Q)\Bigl(1+{11\over 10}
{\alpha_1\over 4\pi}+{9\over 2}{\alpha_2\over 4\pi}\Bigr)-2n_\tg(Q),
&\VII a\cr
b&=102-{38\over 3}n_f(Q)-48n_\tg(Q),&\VII b\cr
c&={2857\over 2}-{5033\over 18}n_f(Q)+{325\over 54}n^2_f(Q).&\VII c\cr}$$
The effects of scalar quarks and Higgs bosons are neglected in accord
with the decoupling theorem.  The off-diagonal two loop effects are
treated as electroweak corrections to the one-loop $a$ coefficient.
We neglect the running of the electroweak corrections, using instead
as an average value
\eqn\VIII{\alpha_{\rm em}=1/133,\quad {\rm sin}^2\theta_W=0.2333,}
and
\eqn\IX{\alpha_1={5\over 3}{\alpha_{\rm em}\over {\rm cos}^2\theta_W},
\quad \alpha_2={\alpha_{\rm em}\over {\rm sin}^2\theta_W}.}
The gluino contribution to the 3-loop $c$ coefficient is not known.
We keep the zero gluino contribution (Eq.~\VII{c}) but monitor the effect
of the $c$ term as an estimate of unknown perturbative effects.
In the fits of reasonable $\chi^2$, the contribution of the $c$ term
is small. We take into account the threshold dependence of quarks and
gluinos according to the formula
\eqna\X
$$\eqalignno{n_f(Q)&=\sum_i^6f\Bigl({Q^2\over 4m_{q_i}^2}\Bigr),&\X a\cr
n_\tg(Q)&=f\Bigl({Q^2\over 4m_\tg^2}\Bigr),&\X b\cr}$$
with \GR\
\eqn\XI{f(x)=1+{1\over 2\sqrt{x(1+x)}}
\ln\Bigl[{{\sqrt{1+x}-\sqrt{x}}\over {\sqrt{1+x}+\sqrt{x}}}\Bigr].}

For each value of $\a3(M_Z)$ and $m_\tg$ we can extrapolate from
the $Z^0$ mass down to the quarkonium region using Eq.~\VI\ and
calculate a $\chi^2$.

The basic assumption given above does not require the total absence of
non-perturbative effects. In fact, in the case of the (most accurately
measured) $1S$ quarkonia states the graphs shown in Ref.~\Lou\ do exhibit
some scatter in the data around the QCD predictions at about the 10\%
level. To find the preferred values of the gluino mass we follow two
alternative customary procedures for the treatment of data in such
cases.  The results of these two procedures do not greatly differ.

\undertext{Method 1:} discard the data point of worst agreement and
minimize the $\chi^2$ of the remaining data points. In the present case
this is the $J/\psi(1S)$ decay. The remaining five vector quarkonia
decays, which contain at least one entry from strange, charm, and bottom
quarks agree well with the theory and provide a relatively sharp minimum
$\chi^2$ as a function of two parameters $\a3(M_Z)$ and the gluino mass
$m_\tg$.

\undertext{Method 2:} retain all six vector quarkonia decays but add in
quadrature with the experimental errors a ``theoretical error" to take
into account possible non-perturbative or binding effects, \ie
\eqn\XII{\a3(\mu_S(i))=\alpha_{3,i}(\mu_S,m_\tg)\pm
\sqrt{\delta_i^2+\lambda_i^2\alpha_{3,i}^2(\mu_S,m_\tg)}.}
On general grounds one would expect such corrections to be more
important for the lighter quarkonia than for the $\Upsilon$ states.
The $\lambda_i$ parameterize our ignorance about higher order and
non-perturbative corrections. They do not, of course, constitute a
model for such effects (since they do not shift the central values of
$\alpha_{3,i}$) and in fact no reliable model exists apart from lattice
QCD which has not as yet attained sufficient numerical accuracy.
Clearly for sufficiently large $\lambda_i$ all predictive power is
lost. We take our basic assumption to imply $\lambda_i\ll 1$. An
adequate $\chi^2$ is found with $\lambda_i=0.05$ for the bottomonium
states, and $\lambda_i=0.10$ for the charmonium and strangeonium vector
states. Our conclusions are qualitatively insensitive to increasing
this value for the lighter quarkonia in the sense that the gluino
mass of minimum $\chi^2$ remains low although the $\chi^2$ values
increase more slowly away from the minimum.
Similarly the favored light gluino also persists for smaller
$\lambda_i$ although the minimum $\chi^2$ is then not a mathematically
acceptable fit.

For comparison with the best fits allowing a light gluino, we show in
Fig.~1 the best fit to the quarkonia data in method 2 assuming the gluino
lies at high mass ($400\GeV$) where it essentially decouples.
This fit seems surprisingly good to the naked eye specially considering
that it relates pure perturbative QCD to the hadronic decay rates of six
vector quarkonia of three species over mass scales varying by a factor
of ten. However it is not a mathematically good fit since it corresponds
to a $\chi^2$ per degree of freedom ($\chi^2/{\rm DoF}$) of $3.7$.
In Fig.~2 we show the
variation in the $\chi^2/6$ for this heavy gluino case as a function
of $\a3(M_Z)$. The minimum $\chi^2$ of 3.7 is several standard deviations
worse than the best fits with a light gluino.  In addition this best fit
corresponds to a value of $\a3(M_Z)$ that is many standard deviations
away from the world average value and is inconsistent with SUSY
unification with a SUSY scale below $10\TeV$. A similar attempt to fit
the five quarkonia states as in method 1 but without light gluinos would
yield a minimum $\chi^2$ per degree of freedom many times larger than
the minimum $\chi^2$ of Fig.~2.

In Fig.~3 we show the $\chi^2$ contours for method 1 treating the gluino
mass $m_\tg$ and $\a3(M_Z)$ as variable.
The $\chi^2/5=1$ contours define two acceptable regions:
\eqna\XIII
$$\eqalignno{\a3(M_Z)&=0.1135\pm 0.0005,\qquad m_\tg=(0.32\pm 0.05) \GeV;
&\XIII a\cr}$$
or
$$\eqalignno{\a3(M_Z)&=0.1145\matrix { +0.0013 \cr -0.0006 \cr},\qquad
m_\tg=\Bigl(0.01\matrix { +0.04 \cr -0.01 \cr}\Bigr) \GeV.
&\XIII b\cr}$$
The best fit corresponding to the central values of \XIII{a}\ is shown in
Fig.~4. Comparing with Fig.~1, it is clear that the fit has improved due
to decay of $\Upsilon$ states into gluino containing hadrons and due to
the slower falloff of the coupling constant as a function of energy.
In the solution corresponding to \XIII{b}, even the $\phi$ decay has
significant contribution from gluino containing final states.
In this case both gluinos must presumably hadronize into a single pion
where they can readily mix with gluon pairs. With light gluinos one must
expect that all hadrons have non-negligible gluino components just as
there is a non-negligible probability to find strange quarks in the sea
of non-strange hadrons.

Fig.~5 shows the $\chi^2$ contours for method 2. In this method the
$\chi^2/6=1$ contour lies within the region:
\eqn\XIV{\a3(M_Z)=0.1115\matrix {+0.0018 \cr -0.0013 \cr},\qquad
         m_\tg=(0.44\pm 0.17) \GeV.}
More conservative values (90\% CL) can be read from the
$\chi^2/{\rm DoF}=2$ contours in Fig.~3 or Fig.~5. In method 2 there
is also a tendency for the $\chi^2$ to drop again toward zero gluino mass
although in this case the $\chi^2/{\rm DoF}$ does not fall below 1
outside of the region of Eq.~\XIV. The $\chi^2/{\rm DoF}<2$ region is
defined by gluino masses less than $1.2\GeV$. The fit to the six vector
quarkonia assuming the central values of Eq.~\XIV\ is shown in Fig.~6.
All three values of $\a3(M_Z)$ picked out by the quarkonia data with
light gluinos are in excellent agreement with the world average value
for this quantity:
\eqn\XV{\a3(M_Z)=0.113\pm 0.003 \qquad  ({\rm World\ Average})\ \HEB.}

There is at present no well established theory of supersymmetry
breaking that would allow the unambiguous prediction of the gluino
mass. Nevertheless, in the most realistic models that have been
extensively studied, supersymmetry is softly broken, triggered
by a super Higgs mechanism in the hidden sector of some minimal
$N=1$ supergravity theories \SUGA.
Therefore, the possible SUSY breaking scenarios
in such models can be parameterized in terms of only a few constants
at the unification scale: the common gaugino mass $m_{1/2}$,
scalar mass $m_0$, and the $A$ and $B$ parameters of dimension mass
characterizing the cubic and quadratic soft-breaking terms that
often exist as well. Although the analysis we presented above
is independent of any specific SUSY breaking models, it is tempting
to discuss the implication of our results to such models.
For simplicity, we now consider such a SUSY GUTs model which assumes the
low energy form of the minimal supersymmetric extension of the standard
model (MSSM) \HK. For our purpose, it is enough to consider three
soft-breaking parameters $m_{1/2}$, $m_0$ and $A$. It is interesting
to note that, in this framework, the low gluino masses favored by
our analysis are natural if the dominant SUSY breaking seed
is the universal scalar mass $m_0$, \ie, $m_{1/2}\ll m_0$.
Such a SUSY breaking pattern has been supported by recent considerations
of proton stability in the context of (non-flipped) $SU(5)$
supergravity \Dick, and also favored by cosmological studies \LYN.
In fact, our results suggest
\eqn\XVI{m_{1/2}=0.}
Such a model might be theoretically appealing since then SUSY
breaking, like electroweak breaking, finds its origin in the
scalar sector. On the other hand, SUSY breaking scenarios with
$m_{1/2}\not =0$ would lead to quite large gluino masses \others.

In the $m_{1/2}=0$ scenario, generally, one expects a supersymmetric
spectrum relatively light compared to what one would get in
other scenarios. Besides the soft-breaking parameters and the currently
unknown top quark mass, one needs two more parameters in order to
specify the full spectrum: the ratio of the two Higgs vev's
${\rm tan}\beta \equiv v_2/v_1$, and the Higgs mixing parameter $\mu$.
In fact, all the gaugino masses then vanish at the tree level and
the gauginos only receive masses through radiative corrections
which are naturally small though dependent upon the masses of other
particles \BGM. In particular, the one-loop corrections
to the gluino and photino (which is an exact mass eigenstate in this
scenario) masses are known, with the dominant contribution coming
from graphs in which the top quark and its two superpartners circulate
around the loop \BGM,
\eqnn\XVII
\eqnn\XVIII
$$\eqalignno{\delta m_\tg&={\a3(m_t)\over 8\pi}m_tF\Bigl(
{m^2_{{\tt}_1}\over m^2_t},
{m^2_{{\tt}_2}\over m^2_t}\Bigr),&\XVII\cr
\delta m_\tp&={\alpha_{\rm em}(m_t)\over 3\pi}m_tF\Bigl(
{m^2_{{\tt}_1}\over m^2_t},
{m^2_{{\tt}_2}\over m^2_t}\Bigr).&\XVIII\cr}$$
where $m_{{\tt}_1,{\tt}_2}$ are the masses of the two scalar top quarks
($m_{{\tt}_1}<m_{{\tt}_2}$), and
\eqn\XF{F(x,y)={\rm sin}2\theta\Bigl[{x\over {1-x}}\ln x
                                     -{y\over {1-y}}\ln y \Bigr].}
The actual gluino and photino masses up to one-loop order
are given by the {\it absolute values} of these mass corrections.
The $\theta$ in \XF\ is the mixing angle between the scalar partners
of left- and right-handed top quarks,
which rotates the ${\tt}_{L,R}$ states into the mass eigenstates
${\tt}_{1,2}$. The overall factor ${\rm sin}2\theta$ was omitted in
Ref.~\BGM, corresponding to the case where the difference between
two diagonal terms of the $2\times 2$ mass-squared matrix of the
scalar top quarks can be neglected (see Eq.~33 below).
The more general result
presented here has been recently calculated by one of us \KY,
and makes transparent the fact that these one-loop mass corrections
vanish exactly if there is no left-right mixing, even if
${\tt}_L$ and ${\tt}_R$ (now mass eigenstates themselves) have
non-degenerate masses. Since we are primarily interested here in the
gluino mass, we neglect in Eq.~\XVIII\ a contribution to the photino
mass from the $W^\pm$-chargino loop diagrams \BGM. A discussion
of this contribution in its general form will be given elsewhere \KY.
As shown below, with currently favored values for the
top quark mass $m_t$ and the two stop quark masses $m_{{\tt}_1}$,
$m_{{\tt}_2}$, the function $F$ is such that gluino masses below
$1\GeV$ are quite natural.

We now discuss the allowed regions for the relevant parameters in
the $m_{1/2}=0$ scenario. To simplify our approach we
neglect the Yukawa couplings for all the fermions although, strictly
speaking, this approximation is only very good for the first two
generations. As a result, the diagonal elements of the sfermion
mass-squared matrix can be written as \KLN\
\eqn\XX{m^2_\tf=m^2_0+m^2_f+M^2_Z {\rm cos}2\beta
 \Bigl[T_{3,f}-e_f{\rm sin}^2\theta_W\Bigr].}
Here we have included the D-term contributions as well.
The average mass-squared of the sfermions is seen to differ from the
average mass-squared of the fermions by the parameter $m^2_0$,
\eqn\XXI{\VEV{m^2_\tf}=m^2_0+\VEV{m^2_f}.}
The average mass-squared of the SUSY partners is approximately equal to
the effective SUSY scale $M^2_S$ that enters into grand unification
considerations. In Ref.~\Lou\ (Eq.~($3.11a$)) it was shown that the
assumption of minimal SUSY unification with a light
gluino (below $M_Z/2$) requires the approximate relation
\eqn\XXII{M_S=150\GeV\times e^{-518.5({\rm sin}^2\theta_W-0.2336)}
e^{1.85(\a3^{-1}(M_Z)-0.113^{-1})}.}
Equating $M^2_S$ with $\VEV{m^2_\tf}$ and substituting the world average
values of ${\rm sin}^2\theta_W$ and $\a3(M_Z)$ from Ref.~\HEB\ yields
within errors the following range for $m_0$,
\eqn\Xa{75\GeV<m_0<270\GeV}
For the top quark mass, we will assume \Ellis\
\eqn\Xb{92\GeV<m_t<147\GeV}
In the $m_{1/2}=0$ scenario, the tree-level masses of the two charginos,
$\chi^\pm_{1,2}$, are given by
\eqn\Xcc{m^2_{\chi^\pm_{1,2}}={1\over 2}\Bigl[2M^2_W+\mu^2\mp \sqrt
{\mu^4+4M^2_W\mu^2+4M^4_W{\rm cos}^22\beta}\Bigr].}
{}From Eq.~\Xcc\ and the requirement that the lightest
chargino ($\chi^\pm_1$) has to be heavier than
about half the $Z^0$ mass \CHGNO, it is found that ${\rm tan}\beta$ is
restricted from both sides, \ie
\eqn\Xc{0.441<{\rm tan}\beta <2.266.}
And for each value of ${\rm tan}\beta$ in the above range there is an
upper limit on the absolute value of the Higgs mixing parameter $\mu$.
Furthermore, the lower limit of $41\GeV$ on the mass of the
light CP-even Higgs boson yields the additional constraints \HIGGS\
\eqn\Xy{0.55<{\rm tan}\beta <0.65;\quad
{\rm or}\quad 1.5<{\rm tan}\beta.}
Combining \Xc\ with \Xy\ yields two allowed ranges for the
${\rm cos}2\beta$ factor of Eq.~\XX,
\eqn\Xz{-0.674<{\rm cos}2\beta <-0.385;\quad
{\rm or}\quad 0.406<{\rm cos}2\beta <0.536.}
The bounds of Eq.~\Xy\ change with the experimental lower limit on
the Higgs mass, becoming inconsistent with Eq.~\Xc\ if this mass is
required to be above $70\GeV$. The allowed parameter space of the
$m_{1/2}=0$ model also requires that the mass of the lightest
chargino be below $M_W$.

The mass-squares of the two scalar top quarks entering into Eqs.~\XVII\
and \XVIII\ are given by
\eqn\Xg{m^2_{{\tt}_1,{\tt}_2}={1\over 2}\Bigl[m^2_{LL}+m^2_{RR}
\mp \sqrt{(m^2_{LL}-m^2_{RR})^2+4m^4_{LR}}\Bigr]}
with the diagonal elements (see Eq.~\XX )
\eqnn\Xd
\eqnn\Xe
$$\eqalignno{m^2_{LL}&=m^2_0+m^2_t+M^2_Z{\rm cos}2\beta\Bigl[
       {1\over 2}-{2\over 3}{\rm sin}^2\theta_W\Bigr],&\Xd\cr
m^2_{RR}&=m^2_0+m^2_t+M^2_Z{\rm cos}2\beta\Bigl[
       {2\over 3}{\rm sin}^2\theta_W\Bigr],&\Xe\cr}$$
and off diagonal element
\eqn\Xf{m^2_{LR}=m_t\Bigl(A_t+{\mu\over {\rm tan}\beta}\Bigr)
\equiv m_tm_0{\tA}_t.}
Here we have introduced the dimensionless mixing parameter
${\tA}_t$ as a useful combination of ${\rm tan}\beta$, $\mu$
and the low energy top soft-breaking parameter $A_t$.
If there is a single source of SUSY breaking corresponding to a
single scale $m_0$, we might expect values of ${\tA}_t$ to be
either zero or of the order unity. In terms of ${\tA}_t$ the
mixing angle factor in Eq.~\XF\ is then
\eqn\Xh{{\rm sin}2\theta={-2m_0m_t{\tA}_t\over
\sqrt{(m^2_{LL}-m^2_{RR})^2+4m^2_0m^2_t{\tA}^2_t}}.}

Imposing the experimental constraint $m_{{\tt}_1}>M_Z/2$ together
with \Xa, \Xb\ and \Xz\ requires that ${\tA}_t<3.2$.
We also use for $\a3$ the value at the top mass
$\a3(m_t)\simeq 0.1$ in evaluating the gluino mass according to
Eq.~\XVII.
In Fig.~7 we show the range of gluino masses predicted by Eq.~\XVII\
as a function of ${\tA}_t$ for the allowed range of values
of ${\tA}_t$, $m_0$, $m_t$, and ${\rm cos}2\beta$ described above.
The constraint from Eq.~\Xg\ that the stop quarks be above half the
$Z^0$ mass is also required in this allowed range.
For each value of ${\tA}_t$ there exist an upper and lower limit
on the gluino mass $m_\tg$. Values of ${\tA}_t$ near zero are
consistent with the near zero gluino masses of Eq.~\XIII{b}.
Values of ${\tA}_t$ near unity are consistent
with the $\chi^2$ minima of Eqs.~\XIII{a}\ and \XIV.
The entire range of values of ${\tA}_t$ assuming Eq.~\Xa\ yields
gluino masses below $1.3\GeV$ in agreement with the quarkonia data
at the two-standard-deviation level. Assuming the result of
Eqs.~\XIII{}\ or \XIV, detailed predictions for the individual squark
and slepton masses within narrow ranges can be made.
Much of the allowed parameter space in Fig.~7 predicts one or more of
the scalar quarks and leptons to have a mass between $M_Z/2$ and $M_Z$.
If the effect of a light gluino and a possible light squark is taken into
account the anomalously large quoted values of $\a3$ coming from the
$Z^0$ hadronic decay can perhaps be reconciled with the world average
value \CC. In addition, using $\alpha (m_t)\simeq 1/127.9$ in
Eq.~\XVIII\ the photino mass $m_\tp$ would then be about five times
smaller than the gluino mass $m_\tg$.
A stable photino of mass about $100\eV$ could provide enough
dark matter to close the universe \EHNOS.
More massive photinos would ``overclose" the universe unless they
could decay efficiently into photon plus gravitino or
annihilate efficiently into photons. Such ultralight
photinos have also been discussed as the explanation of other
astrophysical observations \Sciama.  On the other hand photinos of
such mass have been found to be disfavored \EOSS\ by data from Supernova
1987A unless the squark masses are outside the preferred range
of $60\GeV$ to $2.5\TeV$.
\bigskip
\centerline{CONCLUSIONS}

We have shown that the quarkonia data behave as if there were a gluino
octet in the region below $1\GeV$. Treating the data in either of two
ways, such a light gluino is favored by at least several standard
deviations over the best QCD fits without a light gluino. In addition
the fits without a light gluino are in conflict with SUSY unification
and in disagreement with the world average values of $\a3(M_Z)$ as
discussed in Ref.~\Lou. However, one must remain aware of the usual
possibility that any phenomenological fit to data could be coincidental.
We can not rule out the possibility that the real explanation for the
behavior of the quarkonia data might lie in relativistic binding
corrections or other non-perturbative effects, although this would
contradict the general assumption that the narrowness of the vector
quarkonia is due to perturbative QCD and asymptotic freedom. Confirmation
from other independent data will certainly be required before the results
presented here could be considered compelling. We are presently
investigating the possibility that supporting evidence may be present
elsewhere in existing data or that definitive experimental tests can be
proposed.

In addition we should address the question as to whether such light
gluinos are ruled out by current bounds. The strongest bounds on the
masses of supersymmetric particles come from the decay of the $Z^0$.
Any particle with electroweak charge must lie above about one half the
$Z^0$ mass.

Bounds on other particles such as gluinos or bounds from other processes
are all to a greater extent model dependent. The status of these bounds
is discussed in Ref.~\DataG. Although many lower bounds on the gluino
mass above the mass region indicated here have been quoted, all of these
are to some extent dependent on untested assumptions. For example, the
stringent bounds from the hadron colliders on heavy gluinos allow windows
for light gluinos below $50\GeV$. The low energy windows are well
illustrated in Ref.~\DEQ. Ref.~\DataG\ confirms the lack of unanimous
opinion concerning whether or not very light gluinos have been ruled out.
Many of the purported bounds have obvious loopholes some of which are
pointed out in Ref.~\Lou\ and elsewhere \AEN. For example the CUSB \CUSB\
bound that disfavors gluino masses between $0.6\GeV$ and $2.2\GeV$ from
the non-observation of $\gamma +$gluinoball final states in $\Upsilon$
decay is strongly dependent on the value of the wave function at the
origin of the gluinoball for which only models can be made \KK.
Although the gluino behaves like a quark with a different color charge,
it does have quartic couplings to gluons and gluinos that do not affect
quarkonia in the same way. The binding of gluinos into new hadrons is
therefore more closely related to the binding of gluons into new hadrons
which is a very poorly understood area of hadronic physics at present.
Cosmological constraints on light photinos and gluinos are subject
to similar uncertainties. Most of the range of photino masses between
$100\eV$ and $2\GeV$ is disfavored by one or more cosmological arguments.
However there is a window noted in Ref.~\LK\ for a photino in the mass
range from 4 to $15\MeV$. The gluino would then be in the range from 20
to $75\MeV$ consistent with Eq.~\XIII{b}. It is not clear whether there
is sufficient uncertainty in the cosmological arguments to stretch this
range by a factor of four to accommodate the results of our
Eq.~\XIII{a} or \XIV.

Given the current situation we feel that prudence requires the adoption
of a conservative, non-dogmatic attitude concerning the compatibility
of light gluinos with existing data.

\bigskip
\bigskip
\cl{\bf Acknowledgments}\nobreak
K.Y would like to thank Jorge Lopez for a useful discussion.
This work has been supported in part by the U.S. Department of Energy
under Grant No. DE-FG05-84ER40141 and by the Texas National
Laboratory Research Commission under Grant No. RCFY9155.
\listrefs
\listfigs
\bye